\documentclass[prb,aps,preprint,eqsecnum]{revtex4}
\usepackage[centertags]{amsmath}
\usepackage{amssymb}
\usepackage{amsthm}
\usepackage{graphicx}
\usepackage{epsfig}

\begin{document}

\title{Langevin dynamics with dichotomous noise; direct simulation and applications}

\author{Debashis Barik, Pulak Kumar Ghosh and Deb Shankar Ray{\footnote{
Email address: pcdsr@mahendra.iacs.res.in}}} \affiliation{Indian
Association for the Cultivation of Science, Jadavpur, Kolkata 700
032, India}

\begin{abstract}
We consider the motion of a Brownian particle moving in a potential
field and driven by dichotomous noise with exponential correlation.
Traditionally, the analytic as well as the numerical treatments of
the problem, in general, rely on Fokker-Planck description. We
present a method for direct numerical simulation of dichotomous
noise to solve the Langevin equation. The method is applied to
calculate nonequilibrium fluctuation induced current in a symmetric
periodic potential using asymmetric dichotomous noise and compared
to Fokker-Planck-Master equation based algorithm for a range of
parameter values. Our second application concerns the study of
resonant activation over a fluctuating barrier.
\end{abstract}

\maketitle

\section{Introduction}

Langevin equation \cite{gar,ris,vak} is the key stochastic
differential equation that encompasses a wide range of physical,
chemical and biological sciences. Although originally envisaged as a
descriptor of the motion of a Brownian particle in a fluid in
thermal equilibrium, the equation is now the basic paradigm for
almost any physically realizable stochastic process which takes into
account of the motion of particle in phase space and at the same
time acted upon by a noise force which may be internal or external
depending on whether the fluctuation-dissipation relation is
satisfied or not, respectively. Over the years various noise
processes with arbitrary correlation time have been in use in a wide
variety of problems, \textit{e.g.}, in activated processes in
kinetics \cite{han,cof}, noise-induced transport processes
\cite{mag,rou,doe,milo,jul,rei,but,els,kul,nori,pul,marh} in
condensed matter and biological physics \cite{ser,val,how,tso} etc.
Among others dichotomous or telegraphic noise
\cite{san,kap1,mie,hu,lin,for,eps} plays the typical theoretician's
paradigm of fluctuations in this context. While for linear
potentials the Langevin equation with dichotomous noise is solvable
analytically, arbitrary nonlinear potential poses serious problems.
For example, it is very difficult, if not impossible, to treat a
ratchet device with periodic potential and finite correlation time
of noise. One therefore routinely takes resort to an equivalent
Fokker-Planck approach and the algorithm based on this description.
The notable point is that the approach bypasses the generation of
dichotomous noise as such and a close survey of the recent
literature suggests that a direct simulation of Langevin dynamics of
a particle driven by dichotomous noise still remains to be
addressed. Our object in this paper is thus twofold,

(i) We present a general numerical algorithm for generation of
dichotomous noise with exponential correlation and numerically
simulate the Langevin dynamics for the particle without taking
resort to any Fokker-Planck or master equation description.

(ii) As two prototypical applications with additive and
multiplicative noise processes we consider dichotomous noise-induced
transport or current in a periodic potential in a ratchet device and
resonant activation over a fluctuating barrier.

In implementing the scheme it is essential to take care of the
fact that the numerical error should not bring any additional tilt
to the potential or break of any inversion symmetry or detailed
balance of the system. Second, the forcing terms must be unbiased
so that after appropriate averaging over time, space and ensemble
no directed transport should appear as an artifact. These
considerations are particularly important for thermodynamic
consistency \cite{van,lef,sek,kam,par,tak} of the numerical
scheme. Thus although the numerical simulation of Langevin
dynamics with dichotomous noise is useful in a wide variety of
other situations, our choice of application here is guided by
these considerations.

The outline of the paper is as follows: In Sec.II we present our
method of numerical solution of Langevin equation in presence of an
internal noise and a dichotomous noise. Although the elements of the
present numerical scheme are commonly used in Monte Carlo
calculations, the implementation in generating dichotomous noise in
the numerical solution of Langevin equation for realization of
stochastic path and nonequilibrium fluctuation-induced transport and
resonant activation are new in the current context. Sec.III is
devoted to the numerical results on the noise induced current for a
cosine potential. We compare the methods based on Langevin equation
and Fokker-Planck equation \cite{els} for a range of parameters.
Another application is the study of activated escape over a
fluctuating barrier. The paper is concluded in Sec.IV.

\section{Numerical simulation of Langevin dynamics with dichotomous noise}

In order to motivate our numerical simulation method we consider
the motion of an overdamped particle in a periodic potential,
$V(x)$, simultaneously subjected to an internal thermal noise,
$\xi(t)$, and an external dichotomous or telegraphic noise,
$\eta(t)$. This is described by the Langevin equation of the
following form

\begin{equation}\label{2.1}
\dot{x}(t)=-V^{'}(x)+\xi(t)+\eta(t)
\end{equation}

The periodic potential $V(x)$ may be symmetric or asymmetric
depending on the specificity of the situation. $\xi(t)$ is the
thermal, Gaussian, white noise whose mean and variance are given
by

\begin{eqnarray}
\langle \xi(t) \rangle &=& 0 \label{2.2}\\
\langle \xi(t) \xi(t') \rangle &=& 2 k_B T \delta(t-t')\label{2.3}
\end{eqnarray}

respectively, where $k_B$ is the Boltzmann constant and $T$ is the
absolute temperature which is the measure of the strength of
internal noise.

$\eta(t)$ is the dichotomous noise which can assume only two random
values, say, $a$ and $b$. We also require the random number sequence
to satisfy:

\begin{eqnarray}
\langle \eta(t) \rangle &=& 0 \label{2.4}\\
\langle \eta(t) \eta(t') \rangle &=& \frac{Q}{\tau}
\exp\left(-\;\frac{|t-t'|}{\tau}\right)\label{2.5}
\end{eqnarray}

or in other words , the dichotomous noise must have a zero mean
and be exponentially correlated. $Q$ is the strength and $\tau$ is
the correlation time of the dichotomous noise.

Our problem here is to find out the sample path $x(t)$  of the
particle as a function of time and calculate the relevant
quantities for nonequilibrium fluctuation-induced transport. More
specifically, we calculate the average velocity of the particle
under a steady state condition.

\subsection{The algorithm}

A simple approach for numerical simulation of Eq.(\ref{2.1}) is to
discretize time $t$ and to use a predictor-corrector method in
advancing the particle from $x(t_n)$ to $x(t_{n+1})$ as follows:

\begin{equation}\label{2.6}
x_1(t_{n+1})=x(t_n)-V'(x(t_n)) \Delta t +\eta(t_n) \Delta t + (2
k_B T \Delta t)^{1/2} W_n
\end{equation}

\begin{equation}\label{2.7}
x(t_{n+1})=x(t_n)-\frac{1}{2} \left[ V'(x(t_n)) + V'(x_1(t_{n+1}))
\right]\Delta t + \eta(t_n) \Delta t + (2 k_B T \Delta t)^{1/2} W_n
\end{equation}

where the first step (\ref{2.6}) implies the simplest Euler-type
algorithm for the predictor, whereas the second step corresponds
to the corrector. $\Delta t$ is the time step and $W_n$ is a
Gaussian distributed random number with zero average and a
variance of unity independently chosen at each step using a
standard Box-Muller algorithm \cite{fox,pet,bk,db}.

The next important step is the generation of dichotomous random
number $\eta(t)$ with given properties (\ref{2.4}, \ref{2.5}) such
as, zero average and exponential correlation. To this end we
proceed as follows:

\subsection{Generation of dichotomous noise}

We consider a random variable $\eta(t)$ which switches between two
values $a$ and $b$ randomly in time. The rate of switching from
$a$ to $b$ is $\mu_a$ and from $b$ to $a$ is $\mu_b$. This
two-step process can be described by a probability loss-gain
equation or master equation \cite{gar}

\begin{eqnarray}
\frac{d}{dt}P(a, t|x, t_0) &=& - \mu_a P(a, t|x, t_0) + \mu_b
P(b, t|x, t_0)\label{2.8}\\
\frac{d}{dt}P(b, t|x, t_0) &=& \mu_a P(a, t|x, t_0) - \mu_b P(b,
t|x, t_0)\label{2.9}
\end{eqnarray}

where $P(a, t|x, t_0)$ is the conditional probability that the
variable $\eta(t)$ will assume the value $a$ at some time $t$
given that it was $x$ at earlier time $t_0$. $P(b, t|x, t_0)$ can
be defined similarly.

The conservation of total probability demands

\begin{equation}\label{2.10}
P(a, t|x, t_0)+P(b, t|x, t_0)=1
\end{equation}

The initial condition for Eqs.(\ref{2.8}) and (\ref{2.9}) are
given by:

\begin{equation}\label{2.11}
P(x', t|x, t_0)=\delta_{x' x}\;\;\;\;\;\; at\;\;\;\;\;\; t=t_0
\end{equation}

Eq.(\ref{2.8}) and (\ref{2.9}) admit the following steady state
solutions

\begin{eqnarray}
P^s(a) &=& P(a, \infty|x, t_0)=\frac{\mu_b}{\mu_a+\mu_b}\label{2.12}\\
P^s(b) &=& P(b, \infty|x, t_0)=\frac{\mu_a}{\mu_a+\mu_b}\nonumber
\end{eqnarray}

The conditional probabilities $P(a, t|x, t_0)$ and $P(b, t|x,
t_0)$ for time $t$ can be obtained by solving the Eqs.(\ref{2.8})
and (\ref{2.9}) subject to the conditions (\ref{2.10}),
(\ref{2.11}) and and (\ref{2.12}):

\begin{equation}\label{2.13}
P(a,
t|x,t_0)=\frac{\mu_b}{\mu_a+\mu_b}+\left(\frac{\mu_a}{\mu_a+\mu_b}
\;\delta_{a x}-\frac{\mu_b}{\mu_a+\mu_b}\;\delta_{b
x}\right)\exp\left[-(\mu_a+\mu_b)(t-t_0)\right]
\end{equation}

\begin{equation}\label{2.14}
P(b,
t|x,t_0)=\frac{\mu_a}{\mu_a+\mu_b}-\left(\frac{\mu_a}{\mu_a+\mu_b}
\;\delta_{a x}-\frac{\mu_b}{\mu_a+\mu_b}\;\delta_{b
x}\right)\exp\left[-(\mu_a+\mu_b)(t-t_0)\right]
\end{equation}

Eqs.(\ref{2.13}) and (\ref{2.14}) can be used to calculate the
mean and variance of the dichotomous noise in the steady state as

\begin{equation}\label{2.15}
\langle \eta(t) \rangle = \frac{a\mu_b-b\mu_a}{\mu_a+\mu_b}
\end{equation}

\begin{equation}\label{2.16}
\langle \eta(t) \eta(t') \rangle =
\left(\frac{a\mu_b-b\mu_a}{\mu_a+\mu_b}\right)^2+\frac{\mu_a
\mu_b(a+b)^2}{(\mu_a+\mu_b)^2}\exp\left[-(\mu_a+\mu_b)(t-t')\right]
\end{equation}

respectively. Eqs.(\ref{2.15}) and (\ref{2.16}) can be identified
as dichotomous noise with given properties (\ref{2.4}) and
(\ref{2.5}), respectively, provided

\begin{equation}\label{2.17}
a \mu_b - b \mu_a =0
\end{equation}

with the following identification

\begin{equation}\label{2.18}
\tau=\frac{1}{\mu_a+\mu_b}
\end{equation}

and

\begin{equation}\label{2.19}
\frac{Q}{\tau}=\frac{\mu_a \mu_b(a+b)^2}{(\mu_a+\mu_b)^2}
\end{equation}

Now Eq.(\ref{2.17}) can be rearranged to show $\frac{\mu_a
\mu_b(a+b)^2}{(\mu_a+\mu_b)^2}=a b$, which when used in
(\ref{2.19}) yields

\begin{equation}\label{2.20}
Q=a b \tau
\end{equation}

Furthermore, for convenience, we define an asymmetric parameter

\begin{equation}\label{2.21}
\theta=|a|-|b|
\end{equation}

For choosing a set of seven parameters $a$, $b$, $\mu_a$, $\mu_b$,
$Q$, $\theta$ and $\tau$ it is pertinent to satisfy four relations
(\ref{2.17}), (\ref{2.18}), (\ref{2.20}) and (\ref{2.21}). This
implies that only three independent parameters are needed to
specify the dichotomous noise $\eta(t)$ with zero mean and
exponential correlation.

With these aforesaid preliminaries we are now in a position to
generate realizations of a stochastic process for a dichotomous
noise. This is done in the following way. Let the particle is
located initially (t) at $x_n=a$. To determine whether the
particle moves at time $t_1=t+\Delta t$ to another site
$x_{n+1}=b$ or remain at the same site $x_n=a$, we consider the
following conditional probability as given by (\ref{2.13})

\begin{equation}\label{2.22}
P(a, t_1|a,
t)=\frac{\mu_b}{\mu_a+\mu_b}+\frac{\mu_a}{\mu_a+\mu_b}\exp[-(\mu_a+\mu_b)\Delta
t ]
\end{equation}

An uniformly distributed random number $R$ between $[0, 1]$ is now
generated by the computer. This number is then compared against the
conditional probability (\ref{2.22}). If $P(a, t_1|a, t)>R$, then we
accept the value of the noise $a$, \textit{i.e.}, $x_n=a$, else we
accept the value $b$ or $x_{n+1}=b$. If the value of noise is $b$ at
$t_1(=t+\Delta t)$ then we calculate the conditional probability of
jumping to another site $x_{n+2}=a$ at $t_2(=t_1+\Delta t)$ as

\begin{equation}\label{2.23}
P(a, t_2|b,
t_1)=\frac{\mu_b}{\mu_a+\mu_b}-\frac{\mu_b}{\mu_a+\mu_b}\exp[-(\mu_a+\mu_b)\Delta
t ]
\end{equation}

On the other hand if the value of the noise is $x_{n+1}=a$ at
$t_1(=t+\Delta t)$ we calculate the conditional probability $P(a,
t_2|a, t_1)$ using (\ref{2.22}). We then compare the probability
$P(a, t_2|b, t_1)$ or $P(a, t_2|a, t_1)$ against another uniformly
distributed random number $R_1$ between [0, 1].

If $P(a, t_2|b, t_1)$ or $P(a, t_2|a, t_1) > R_1$ then the value
of the noise at $t_2$ is $x_{n+2}=a$ else we accept $x_{n+2}=b$.
By repeating the procedure we can generate a sequence of random
numbers $\eta(t)$ switching between two values $a$ and $b$.

It is important to note that the time interval between the two
steps is always fixed and is equal to $\Delta t$ which is much
smaller than the correlation time ($\tau$) of the noise ($\Delta t
\ll \tau$).

Fig.1(a-b) shows two illustrative dichotomous noise profiles for
several values of $\tau$. A close look into the figures suggests
that with increase of correlation time, residence time of a
particular state increases on an average. In order to check the
numerical accuracy of the method of generation of dichotomous noise
$\eta(t)$ we first calculate its first moment $\langle \eta(t)
\rangle$ satisfying the condition (\ref{2.17}). The typical ensemble
averaging is carried out over long time of $100 \times 5000$ time
steps as well as over $1000$ sample paths. The ensemble average
$\langle \eta(t) \rangle$ is close to zero (less than $\sim 10^{-6}$
as determined numerically). In Fig.2 we calculate the normalized
auto-correlation function $\frac{\langle
\eta(t)\eta(t+t')\rangle}{\langle\eta(t)^2\rangle}$ of the noise for
$a=6.0$, $b=-4.0$ for several given values of $\tau (0.5, 2.0\;$ and
$5.0)$. The values of $\tau$ are compared to the correlation time
obtained by the corresponding numerical fitting curves. The
agreement is found to be satisfactory.

\section{Results: (A) Application to noise induced current}

Having explored the method of generation of dichotomous noise we
now proceed to solve the Langevin equation (\ref{2.1}) or its
discretized version (\ref{2.6}) and (\ref{2.7}). The typical
sample paths for the particle for several values of noise strength
$Q$ are plotted in Fig.3. It is apparent that while for symmetric
noise ($\theta = 0$) the system does not feel any additional load,
asymmetry in the dichotomous noise gives rise to an external tilt
depending on the strength of noise. In what follows we note that
this asymmetry induced tilt in the overall potential is
particularly important for generation of current when the
potential is symmetric. The important relevant quantity for our
present study is the steady state average velocity which can be
defined as

\begin{equation}\label{3.1}
\langle v \rangle=\frac{1}{S}\sum_S
\left(\frac{1}{\cal{T}}\;\frac{1}{N}\sum_N
\left(x_N-x(0)\right)\right)_S
\end{equation}

$\cal{T}$ is the total time over which the displacements are
averaged over for a particular sample path. The second averaging
is over the number of sample paths, $S$. Ensemble averaging
requires both $\cal{T}$ and $S$ to be sufficiently large.

We now consider a cosine potential, \textit{i.e.}, $V(x)=\cos(x)$,
a symmetric periodic potential with periodicity $2\pi$. For the
entire simulation work we have chosen the time step of integration
$\Delta t=0.01$, number of time step is typically of the order of
$10^6$ and the number of sample paths is around $1000$. We have
calculated the steady state average velocity as a function of
correlation time $\tau$ of the dichotomous noise keeping the
asymmetric parameter $\theta=0$ and shown $\langle v \rangle$ vs.
$\tau$ profile in Fig.4. As expected we observe that the average
velocity is zero, since the potential as well as the external
dichotomous noise are symmetric and there can not be any directed
motion of the particle. Fig.4 thus serves as a thermodynamic
consistency check for our numerical calculation.

In order to exhibit Brownian ratchet effect we now switch on the
asymmetric parameter $\theta$ to a non-zero value so that $\theta
\neq 0$. In Fig.5 we display the variation of current $\langle v
\rangle$ as a function of correlation time $\tau$ for three
different values of $\theta(=-1,-2,-4)$. As the magnitude of
$\theta$ increases the peak value of the current increases
significantly. When $\tau$ is small $\langle v \rangle$ tends to
be vanishing. This is because in order to have a net directed
motion it is required that the system must not relax to its
equilibrium state instantaneously. On the other hand when $\tau$
is very large, the overdamped system gets too much time to reach
its equilibrium so that on an average net flows in the left and
right directions tend to equalize making a net drift current
vanishingly small. The system behaves resonantly at some optimum
value of correlation time $\tau$ of the dichotomous noise, for
which the current is maximum.

In Fig.6(a-b) we exhibit the variation of current ($\langle v
\rangle$) as a function of correlation time $\tau$ of the
dichotomous noise for two different values of asymmetry parameter
$\theta$ given by solid lines using Langevin dynamics. The results
are compared with those obtained using Fokker-Planck-Master equation
\cite{els} with periodic boundary condition. In Fig.7(a-b) we make a
comparison between $\langle v \rangle$ vs. $Q$ profile (for two
different values of $\tau$) calculated by our Langevin method and
the Fokker-Planck method. The agreement is found to be quite
satisfactory and provides support to our simulation method based on
Langevin dynamics.

{\bf (B) RESONANT ACTIVATION OVER A FLUCTUATING BARRIER}

Our second application of the method concerns resonant activation -
the well-known resonance effect
\cite{dor,gam,rein,chin,bier,van1,mac,pkg} which can be observed in
the variation of mean first passage time as a function of flipping
rate of fluctuation of the barrier height of a double well
potential. For the present purpose we consider a situation where the
fluctuation is dichotomic in nature, \textit{i.e.}, the barrier
height fluctuates between two values. The fluctuation in potential
has been found to be important in several kinetic models
\cite{bier,van1} for chemical reactions. We consider the potential
of the form

\begin{equation}\label{3.2}
V(x,t)=U(x)+\frac{1}{2} x^2 \eta(t)
\end{equation}

where $U(x)$ is a bistable potential
($-\frac{A}{2}x^2+\frac{B}{4}x^4$, $A$ and $B$ being constants) with
a barrier at metastable point $x=0$ and two stable minima of the
potential are at $x=\pm \sqrt{\frac{A}{B}}$. The fluctuations in the
barrier height of the potential are due to dichotomous noise,
$\eta(t)=\{a,b\}$, \textit{i.e.}, the barrier height of the
potential fluctuates between two states randomly. Under overdamped
condition the reaction coordinate $x(t)$ is governed by the
following Langevin equation

\begin{equation}\label{3.3}
\gamma \dot{x}=A x - B x^3 + x \eta(t) + \xi(t)
\end{equation}

$\xi(t)$ is the usual thermal internal noise as defined in
Eq.(\ref{2.2}) and Eq.(\ref{2.3}) An important point of departure
from the earlier equation (\ref{2.1}) is the presence of a
multiplicative noise (in the third term on the right hand side of
Eq.(\ref{3.3})) which is Gaussian distributed and exponentially
correlated. Since, the stochastic integrals are different in
Statonovich and \^{I}to definitions, the drift term corresponding to
Langevin equation (\ref{3.3}) is $A x - B x^3$ according to
\^{I}to's prescription, whereas Statonovich prescription provides a
drift term of the form $A x - B x^3+x$. The spurious drift component
however is missing in \^{I}to's definition. It is therefore somewhat
convenient to use \^{I}to's scheme for direct simulation of Langevin
equation. In what follows we adopt this scheme and use of the
predictor-corrector steps as (with $\gamma=1.0$);

\begin{equation}\label{3.4}
x_1(t_{n+1})=x(t_n)+[Ax(t_n)-Bx^3(t_n)+x(t_n)\eta(t_n)]\Delta t + (2
k_B T \Delta t)^{1/2} W_n
\end{equation}

\begin{eqnarray}\label{3.5}
x(t_{n+1})=x(t_n)&+&\frac{1}{2}
[\{Ax(t_n)-Bx^3(t_n)+x(t_n)\eta(t_n)\}\nonumber\\
&+&\{Ax_1(t_n)-Bx_1^3(t_n)+x_1(t_n)\eta(t_n)\}]\Delta t + (2 k_B T
\Delta t)^{1/2} W_n
\end{eqnarray}

where the averages and variances of the white and dichotomous noises
are as given earlier in the context of Eq.(\ref{2.1})

To analyze the essential features of the activated escape following
stochastic dynamics over a fluctuating potential barrier we
numerically simulate the Langevin equation (\ref{3.4}-\ref{3.5})
using the method for dichotomous noise generation as developed in
the preceding section with a very small time time step ($\Delta
t=0.01$). In our simulation we follow the dynamics of the particle
starting from the potential minimum at $x=- \sqrt{\frac{A}{B}}$ of
the left well till it arrives at the barrier top at $x=0$ where the
particle is removed. The first passage time \cite{str} being a
statistical quantity due to random nature of the dynamics we
calculate the statistical average of the first passage time over
$2000$ trajectories. We chose the parameters of the potential as
$A=0.5$ and $B=0.1$ for the entire set of calculation. We present
our simulation results in Fig.8 and Fig.9 for several values of
temperatures and variance ($\frac{Q}{\tau}$) of dichotomous noise.
From both the figures it is clear that the mean first passage time
($\langle T \rangle$) passes through a minimum as one increases the
correlation time ($\tau$) of dichotomic barrier fluctuation. It is
apparent that the mean first passage time responds resonantly with
the correlated fluctuations of the barrier height. As one increases
the strength of the internal noise of heat bath the escape rate
increases (or the mean first passage time decreases) and the
resonance behavior is manifested at the very high flipping rate as
evident from Fig.8. When the flipping rate of the barrier height is
very high ($\frac{1}{\tau} \rightarrow 0$) the system effectively
feels an average barrier height so the mean first passage time is
independent of the flipping rate. This is evident form the inset of
Fig.8. Fig.8 and Fig.9 show that the mean first passage time becomes
almost independent of the correlation time in the limit of high
correlation time or slow flipping rate.
\section{Conclusion}

We have presented an algorithm for numerical simulation for
generating dichotomous noise and solution of the Langevin equation.
As immediate applications of the method we have calculated
nonequilibrium fluctuation induced current due to asymmetric,
exponentially correlated dichotomous noise in a symmetric periodic
potential over a wide range of parameter values and resonant
activation rate of barrier crossing in a double well potential. The
method is compared to the Fokker-Planck equation based numerical
technique for solving stochastic dynamics on a discrete lattice
simulated by a Master equation and is found to be complementary and
well-suited for calculation of ratchet effect. We emphasize that
while the overwhelming majority of the treatment of various
phenomena including barrier crossing dynamics, resonant activation,
stochastic resonance using dichotomous noise rely heavily on
Fokker-Planck-Master equation approach we anticipate that the
present method of solution of Langevin dynamics by direct simulation
of dichotomous noise will be useful for various purposes in these
issues. The simulation method can be easily extended to other noise
processes involving, for example, three state jump process and for a
wide range of correlation time and potentials without much
difficulty.

{\bf Acknowledgement:}\\
Thanks are due to the Council of Scientific and Industrial Research
for partial financial support.

\newpage

\begin{center}
{\bf Figure Captions}
\end{center}

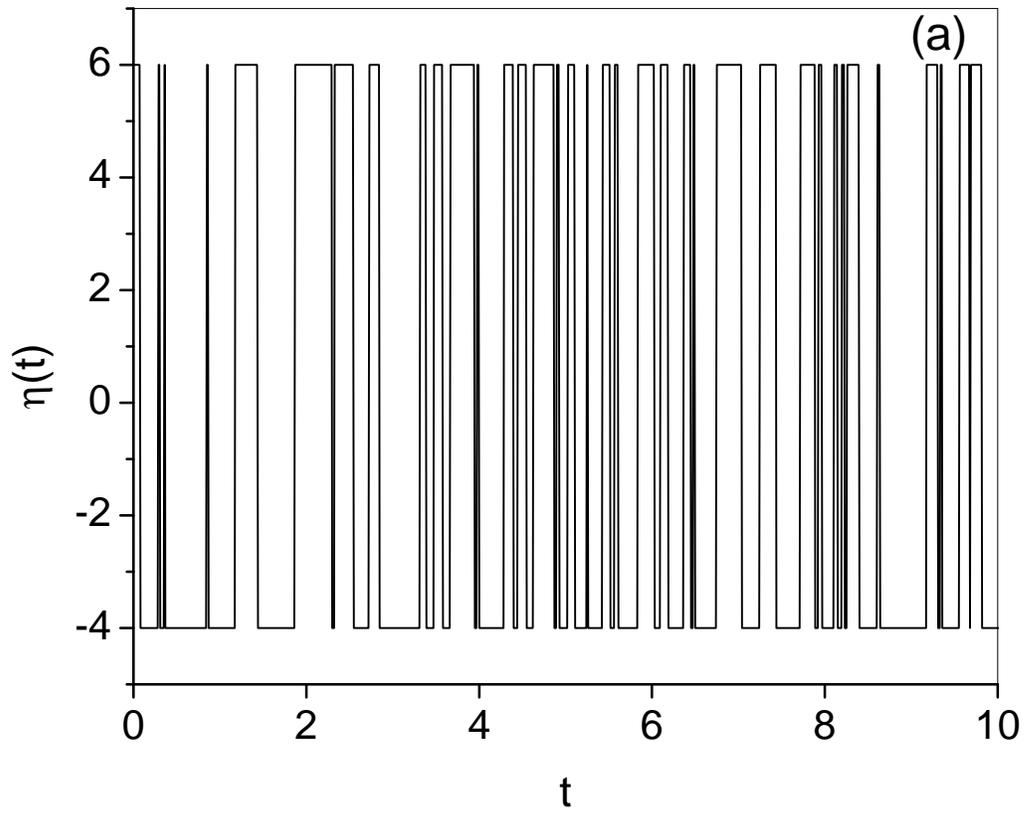
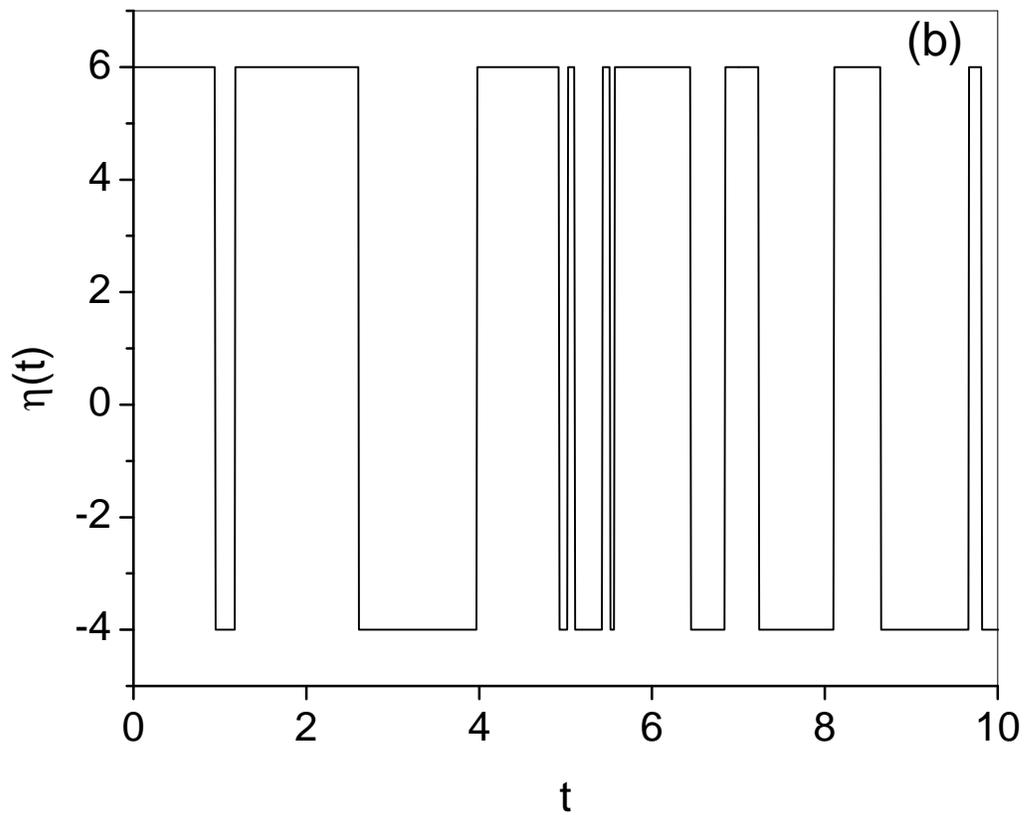
Fig.1: Generation of dichotomous noise profile for $a=6$, $b=-4$
(a) $\tau = 0.05$ and (b) $\tau = 0.4$.

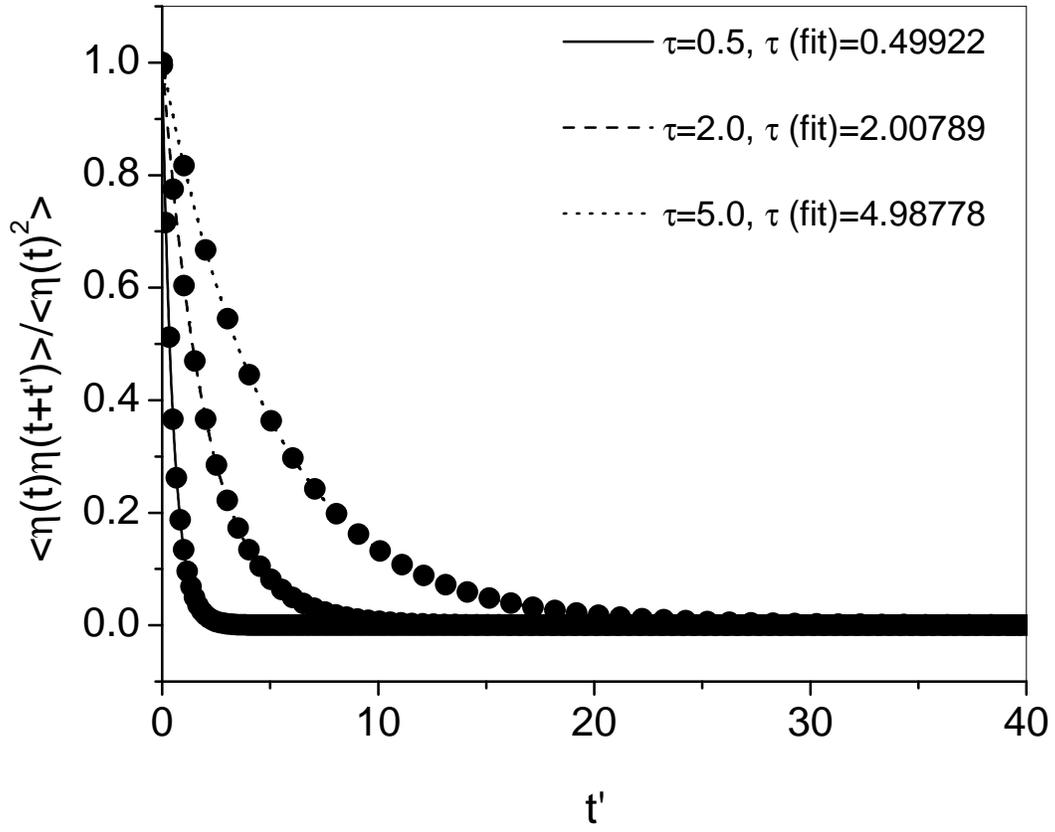
Fig.2: Plots of normalized correlation function vs. $t'$ for
$a=6.0$, $b=-4.0$, for three different noise profiles with given
correlation times $\tau=0.5, 2.0$ and $5.0$. Bold circles
corresponds to fitted curves.

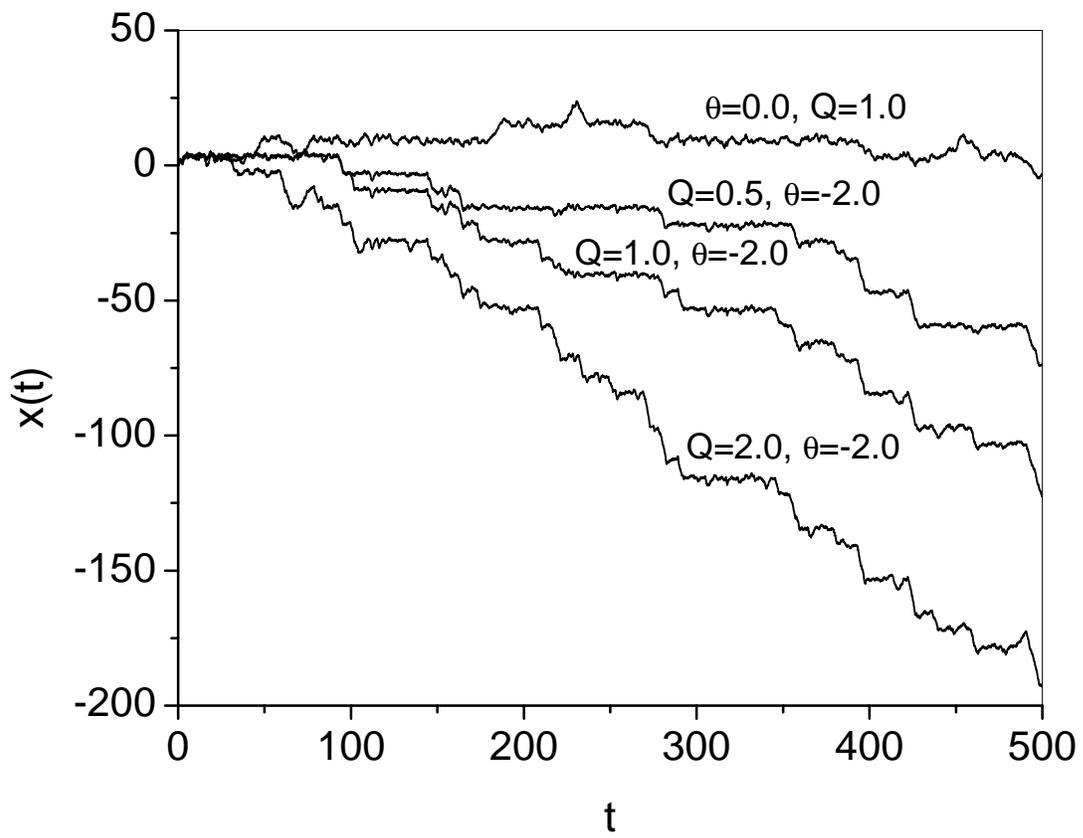
Fig.3: Plot of sample paths of the particle against time for
different dichotomous noise strength ($Q$) and asymmetry parameter
($\theta$) with $D=0.2$ and $\tau=1.0$.

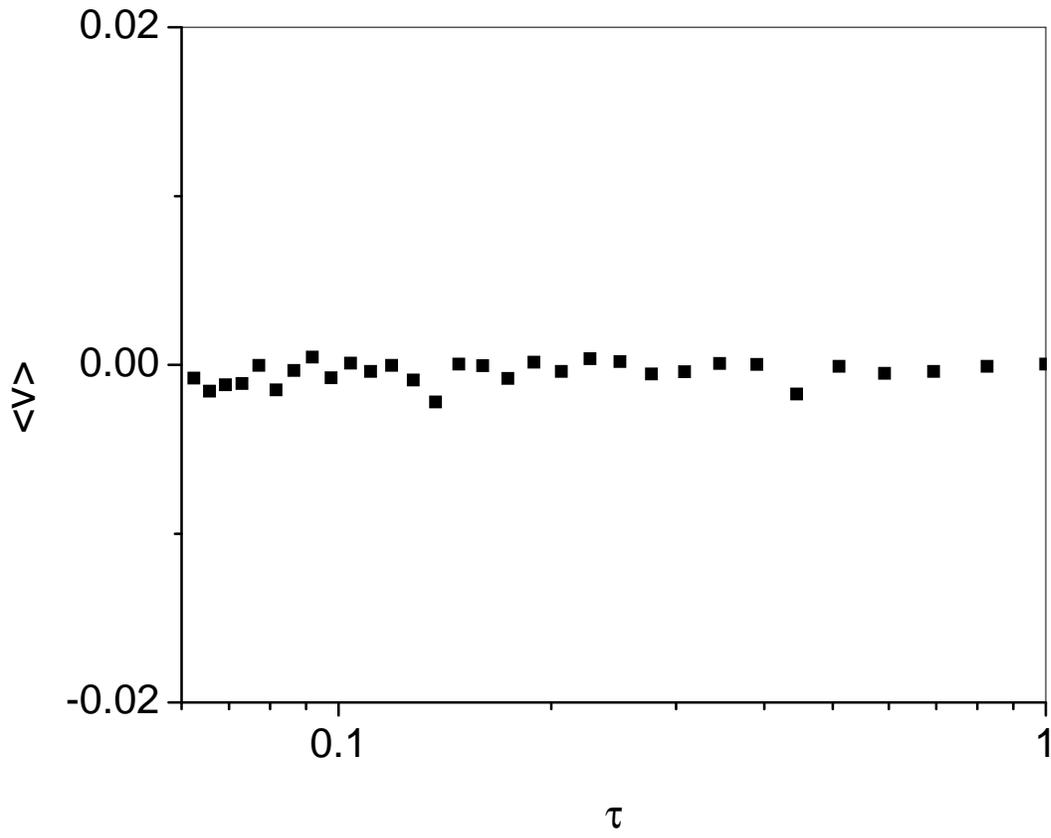
Fig.4: Plot of average velocity $\langle v \rangle$ with
correlation time $\tau$ of dichotomous noise for asymmetry
parameter $\theta=0$, $D=0.02$ and $Q=1.0$.

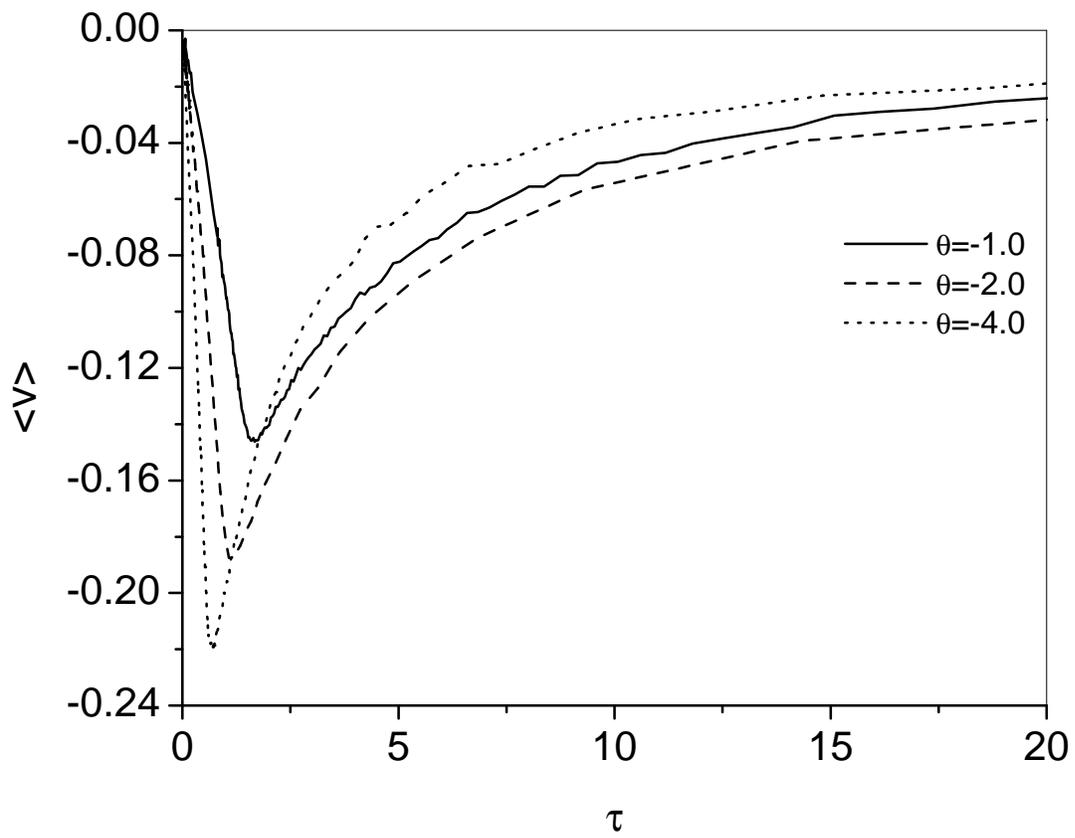
Fig.5: Plot of $\langle v \rangle$ vs. $\tau$ with different
values of $\theta$ for constant white noise strength $D=0.02$ for
constant dichotomous noise strength $Q=3.0$.

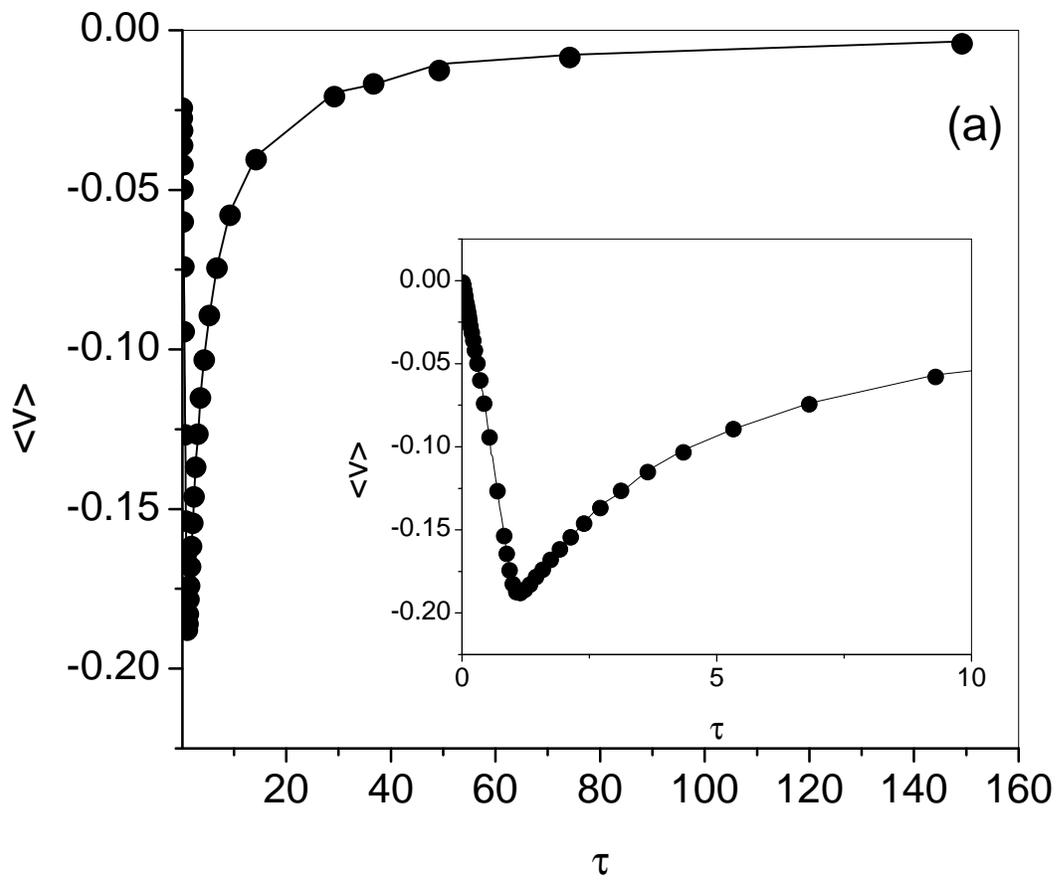
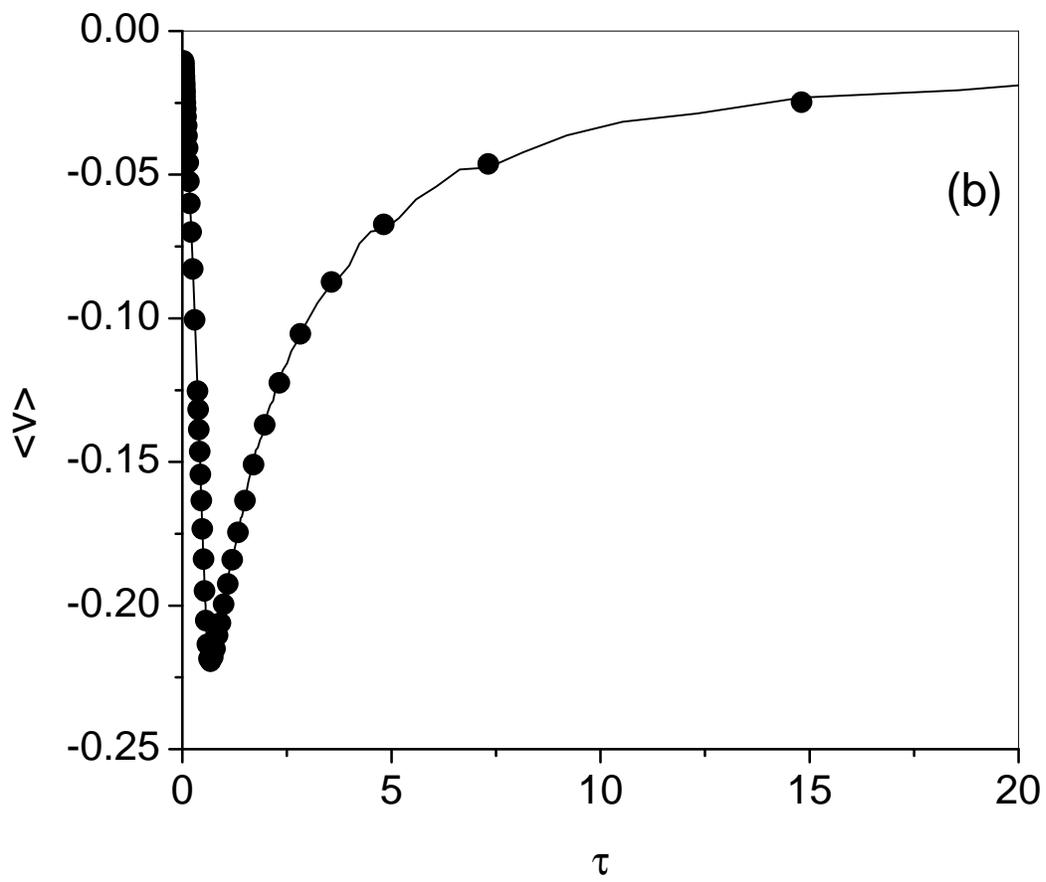
Fig.6: (a) Average velocity ($\langle v \rangle$) vs. dichotomous
noise correlation time ($\tau$) profile is compared with
Fokker-Planck-Master equation method (solid circle) for
$\theta=-2.0$, $Q=3.0$ and $D=0.02$. Inset: Same plot but on a
shorter time scale. (b) Same as in Fig.6(a) but for
$\theta=-4.0$.

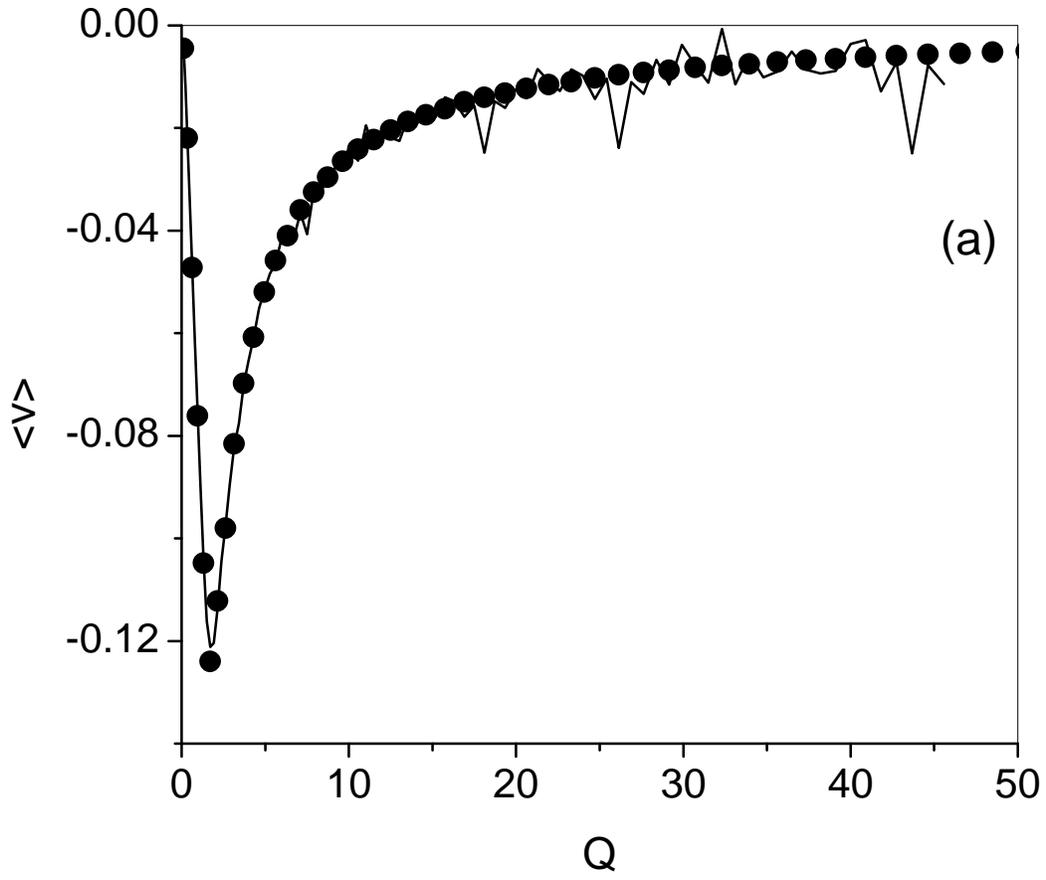
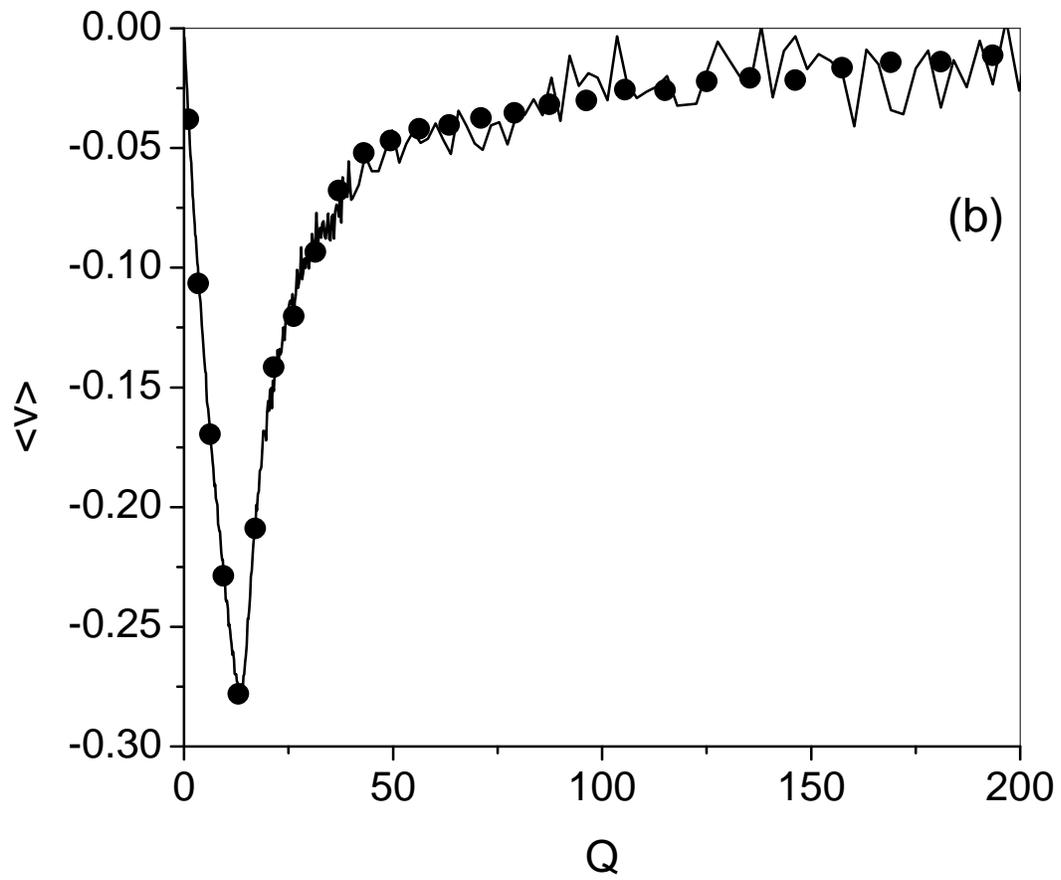
Fig.7: (a) Average velocity ($\langle v \rangle$) vs. dichotomous
noise strength ($Q$) profile is compared with Fokker-Planck-Master
equation method (solid circle) for $\tau=0.5$, $\theta=-2.0$ and
$D=0.02$. (b) Same as in Fig.7(a) but for $\tau=5.0$.

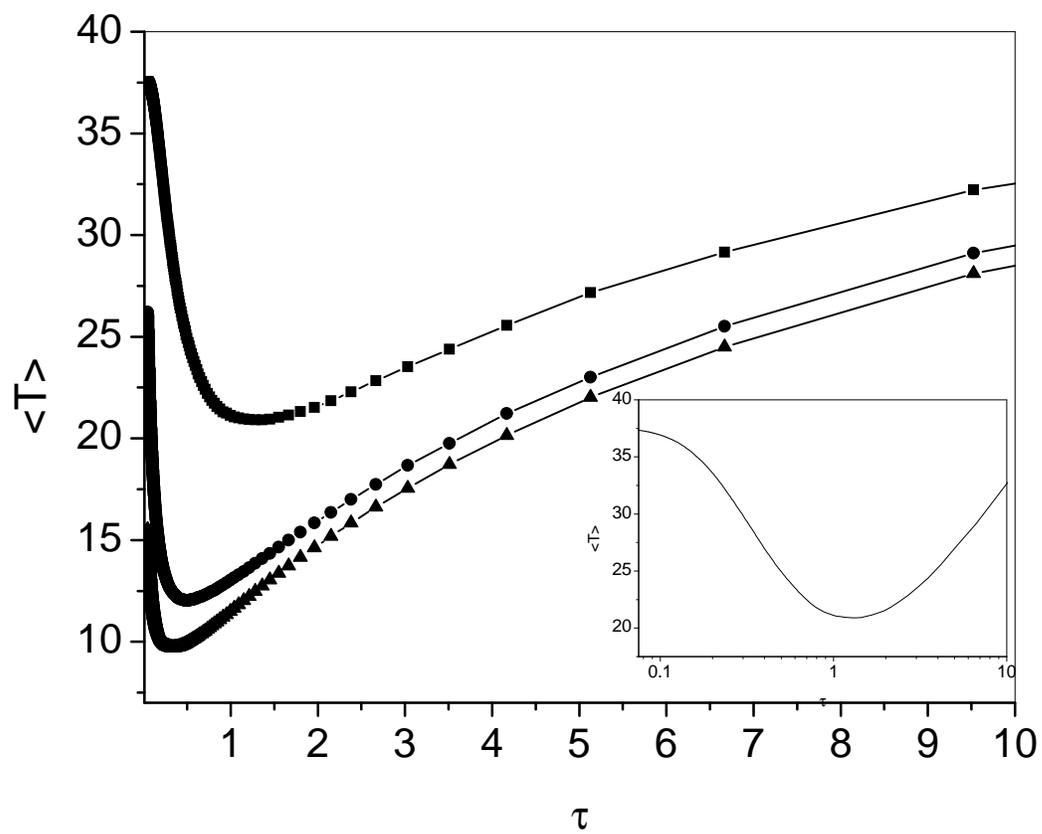
Fig.8: Mean first passage time ($\langle T \rangle$) vs. correlation
time of dichotomous noise $\tau$ for $D=0.1$ (box); $D=1.5$ (circle)
and $D=3.0$ (triangle). (The inset:mean first passage time ($\langle
T \rangle$) vs. correlation time of dichotomous noise $\tau$ on a
log scale for $D=0.1$) for $a=2.0$ and $b=1.0$.

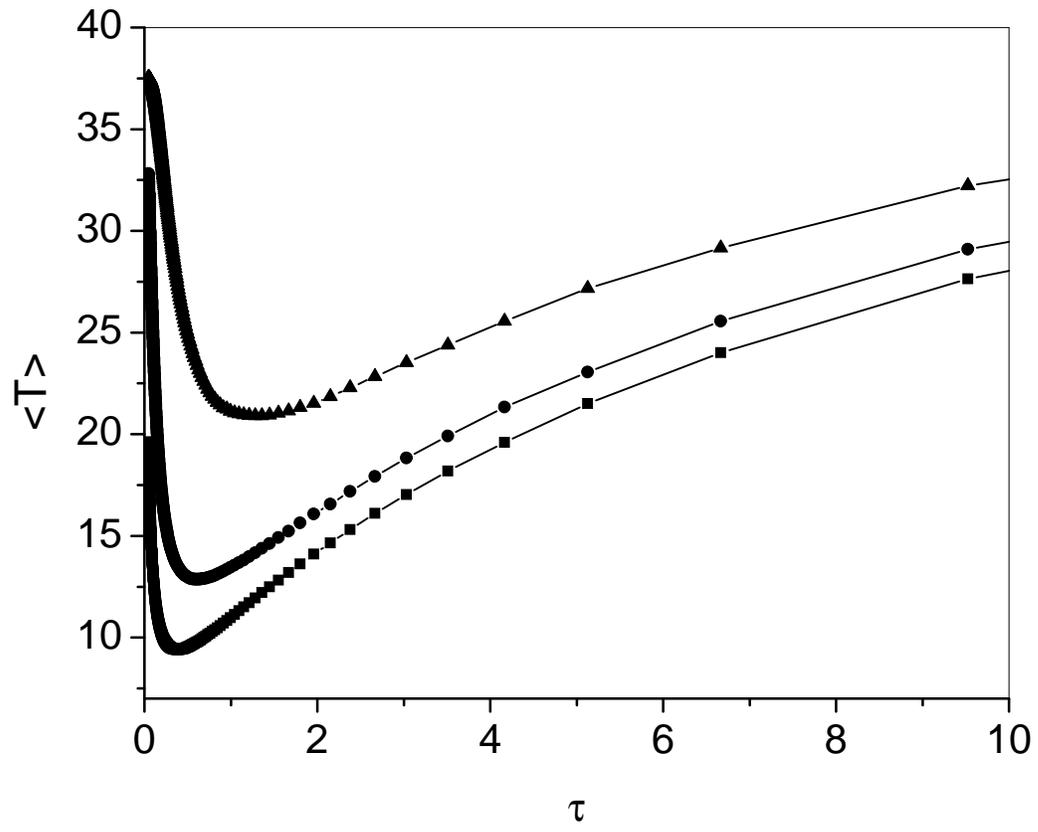
Fig.9: Mean first passage time ($\langle T \rangle$) vs. correlation
time of dichotomous noise $\tau$ for several values of variances of
dichotomous noise [$a=3.0$, $b=2.0$ (triangle); $a=2.5$, $b=1.5$
(circle) and $a=2.0$, $b=1.0$ (square)] for $D=0.1$.
\end{document}